# VLSI Implementation of Cascaded Integrator Comb Filters for DSP Applications

Rozita Teymourzadeh , Masuri Othman

Department of Electrical, Electronic and Systems Engineering
VLSI Design Research Group
National University of Malaysia
rozita60@vlsi.eng.ukm.my

**Abstract**

*The recursive comb filters or Cascaded Integrator Comb filter (CIC) are commonly used as decimators for the sigma delta modulators. This paper presents the VLSI implementation, analysis and design of high speed CIC filters which are based on a low-pass filter. These filters are used in the signal decimation which has the effect on reducing the sampling rate. It is also chosen because its attractive property of both low power and low complexity since it dose not required a multiplier. Simulink toolbox available in Matlab software which is used to simulator and Verilog HDL coding help to verify the functionality of the CIC filters. Design procedures and examples are given for CIC filter with emphasis on frequency response, transfer function and register width. The implementation results show using Modified Carry Look-ahead Adder for summation and also apply pipelined filter structure enhanced high speed and make it more compatible for DSP applications.*

## 1 Introduction

Electronic and communication system for speech processing and radar make use of sigma delta modulator in their operation [1], [2]. Future systems are required to operate with high speed and therefore the sigma delta modulator must be designed accordingly.
Sigma delta ($\sum\Delta$) modulator is an over sampled modulation technique which provides high resolution sample output in contrast to the standard Nyquist sampling technique. However at the output, the sampling process is needed in order to bring down the high sampling frequency and obtain high resolution. The CIC filter is a preferred technique for this purpose. In 1981, Eugene Hogenauer [3] invented a new class of economical digital filter for decimation called a Cascaded Integrator Comb filter (CIC) or recursive comb filter. Additionally the CIC filter does not require storage for filter coefficients and multipliers as all coefficients are unity [4]. Furthermore its on-chip implementation is efficient because of its regular structure consisting of two basic building blocks, minimum external control and less complicated local timing is required and its change factors is reconfigurable with the addition of a scaling circuit and minimal changes to the filter timing. It is also used to perform filtering of the out of band quantization noise and prevent excess aliasing introduced during sampling rate decreasing. Hence enhanced high speed will be key issue in chip implementation of CIC decimators. This filter consists of three parts which are Integrator, comb and down sampler. CIC filter is considered as recursive filter because of the feedback loop in integrator circuit.

The next section describes the mathematical formulation and block diagram of CIC filters in detail. Enhanced high speed architecture is explained in section 3. Section 4 shows implementation and design result in brief. Finally conclusion is expressed in section 5.

## 2 An overview of decimation system

The purpose of the CIC filter is twofold; firstly to remove filtering noise which could be aliased back to the base band signals and secondly to convert high sample rate m-bit data stream at the output of the Sigma-delta modulator to n-bit data stream with lower sample rate. This process is also known as decimation which is essentially performing the averaging and a rate reduction functions simultaneously. Figure 1 shows the decimation process using CIC filter.

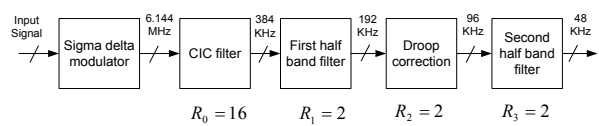

*Fig. 1 Digital Decimation Process*

The two half band filters [7] are used to reduce remain sampling rate reduction to the Nyquist output rate. First half band filter and second half band filter make the frequency response more flat and sharp similar to ideal filter frequency response.

Droop correction filter is allocated to compensate pass band attenuation which is created by CIC filter. The frequency response of overall system will be shown in section 4.

## 2.1 CIC filters structure

The CIC filter consist of N stages of integrator and comb filter which are connected by a down sampler stage as shown in figure 1 in z domain. The CIC filter has the following transfer function:

$$H(z) = H_I^N(z).H_C^N(z) = \frac{(1-z^{-RM})^N}{(1-z^{-1})^N} = (\sum_{k=0}^{RM-1} z^{-k})^N \quad (1)$$

where N is the number of stage, M is the differential delay and R is the decimation factor.
In this paper, N, M and R have been chosen to be 5, 1 and 16 respectively to avoid overflow in each stages.

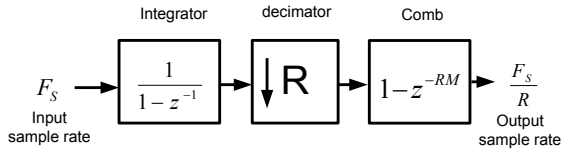

*Fig. 2 One-stage of CIC filter block diagram*

N, M and R are parameters to determine the register length requirements necessary to assure no data loss. Equation (1) can be express as follow:

$$H(z) = \sum_{k=0}^{(RM-1)N} h(k)z^{-k} = \left[\sum_{k=0}^{RM-1} z^{-k}\right]^N \leq \left|\sum_{k=0}^{RM-1} z^{-k}\right|^N$$

$$\leq \left(\sum_{k=0}^{RM-1} |z|^{-k}\right)^N = \left(\sum_{k=0}^{RM-1} 1\right)^N = (RM)^N \quad (2)$$

From the equation, the maximum register growth/width, $G_{max}$ can be expressed as:

$$G_{max} = (RM)^N \quad (3)$$

In other word, $G_{max}$ is the maximum register growth and a function of the maximum output magnitude due to the worst possible input conditions [3].
If the input data word length is $B_{in}$, most significant bit (MSB) at the filter output, $B_{max}$ is given by:

$$B_{max} = [N\log_2 R + B_{in} - 1] \quad (4)$$

In order to reduce the data loss, normally the first stage of the CIC filter has maximum number of bit compared to the other stages. Since the integrator stage works at the highest oversampling rate with a large internal word length, decimation ratio and filter order increase which result in more power consumption and speed limitation.

## 2.2 Truncation for low power & high speed purpose

Truncation means estimating and removing Least Significant Bit (LSB) to reduce the area requirements on chip and power consumption and also increase speed of calculation. Although this estimation and removing introduces additional error, the error can be made small enough to be acceptable for DSP applications.
Figure 2 illustrates five stages of the CIC filter when $B_{max}$ is 25 bit so truncation is applied to reduce register width. Matlab software helps to find word length in integrator and comb section.

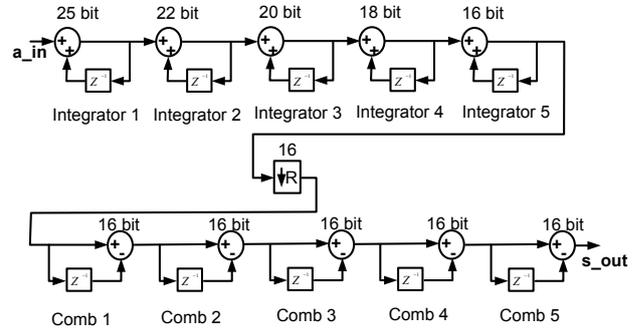

*Fig. 3 Five-stages of truncated CIC filter include integrator and comb cell*

## 3 Speed Improvement

There are two ways to speed up the CIC filter which describes as below.

### 3.1 Pipeline structure

One way to have high speed CIC filter is by implementing the pipeline filter structure. Figure 4 shows pipeline CIC filter structure when truncation is also applied. In the pipelined structure, no additional pipeline registers are used. So that hardware requirement is the same as in the non-pipeline [6]. CIC decimation filter clock rate is determined by the first integrator stage that causes more propagation delay than any other stage due to maximum number of bit. So it is possible to use a higher clock rate for a CIC decimation filter if a pipeline structure is used in the integrator stages, as compared to non-pipelined integrator stages. Clock rate in integrator section is R times higher than in the comb section, so pipeline structure can not applied for comb section.

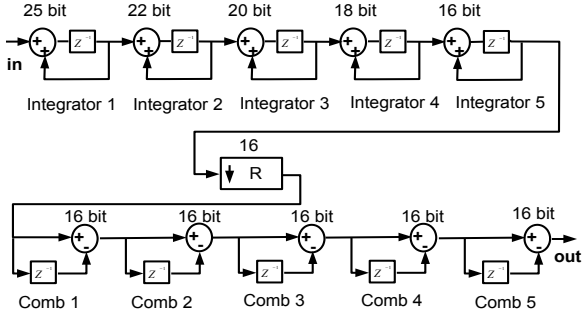

**Fig. 4** *Five-stage of truncated pipeline CIC filter include integrator and comb cell*

### 3.2 Modified Carry look-ahead Adder (MCLA)

The other technique to increase speed is using Modified Carry Look-ahead Adder. The Carry Look-ahead adder (CLA) is the fastest adder which can be used for speeding up purpose but the disadvantage of the CLA adder is that the carry logic is getting quite complicated for more than 4 bits so Modified Carry Look-ahead Adder (MCLA) is introduced to replace as adder. This improve in speed is due to the carry calculation in MCLA. In the ripple carry adder, most significant bit addition has to wait for the carry to ripple through from the least significant bit addition. Therefore the carry of MCLA adder has become a focus of study in speeding up the adder circuits [5]. The 8 bit MCLA structure is shown in Figure 6. Its block diagram consists of 2, 4-bit module which is connected and each previous 4 bit calculates carry out for the next carry. The CIC filter in this paper has five MCLA in integrator parts. The maximum number of bit is 25 and it is decreased in next stages. So it truncated respectively to 25, 22, 20, 18 and 16 bit in each adder, left to right Notice that each 4-bit adder provides a group Propagate and Generate Signal, which is used by the MCLA Logic block. The group Propagate $P_G$ and Generate $G_G$ of a 4-bit adder will have the following expressions:

$$P_G = p_3.p_2.p_1.p_0 \quad (5)$$
$$G_G = g_3 + p_3.g_2 + p_3.p_2.g_1 + p_3.p_2.p_1.g_0 \quad (6)$$

The most important equations to obtain carry of each stage have been defined as below:

$$c_1 = g_0 + (p_0.c_0) \quad (7)$$
$$c_2 = g_1 + (p_1.g_0) + (p_1.p_0.c_0) \quad (8)$$
$$c_3 = g_2 + (p_2.g_1) + (p_2.p_1.g_0) + (p_2.p_1.p_0.c_0) \quad (9)$$
$$c_4 = g_3 + (p_3.g_2) + (p_3.p_2.g_1) + (p_3.p_2.p_1.g_0) + (p_3.p_2.p_1.p_0.c_0) \quad (10)$$

Calculation of MCLA is based on above equations. 8-Bit MCLA Adder could be constructed continuing along in the same logic pattern, with the MSB carry-out resulting from OR & AND gates.

The Verilog code has been written to implement summation. The MCLA Verilog code was downloaded to the Xilinx FPGA chip.

It was found minimum clock period on FPGA board is 4.389ns (Maximum Frequency is 220 MHz).

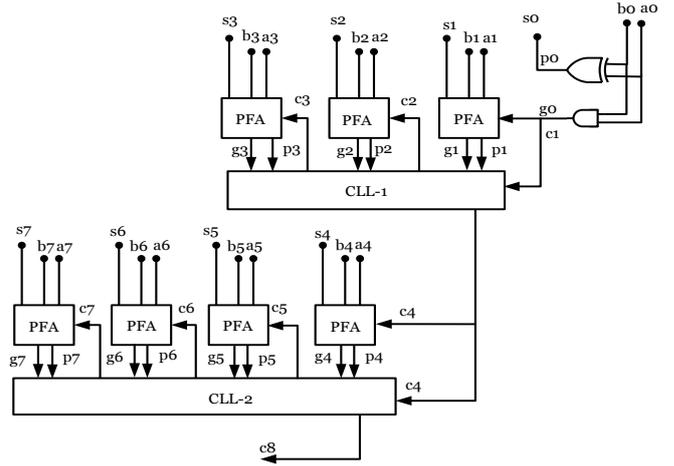

**Fig. 5** *The 8 bit MCLA structure*

### 4 Design Results

Figure 6 illustrate the frequency response of the CIC filter when the sampling frequency is 6.144 MHz and the pass band frequency is 348 KHz.

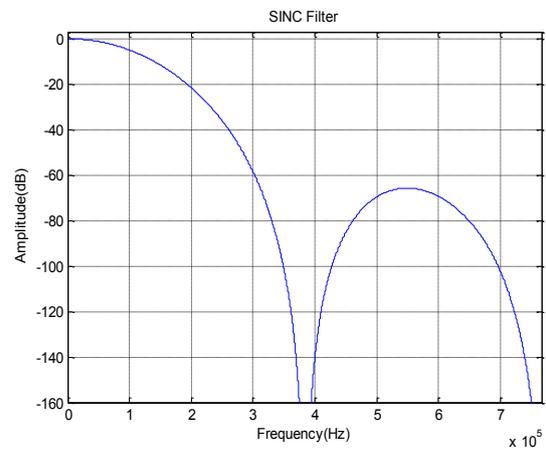

**Fig. 6** *Comb filter frequency response for R=16, M=1 and N=5*

After the sigma delta modulator, the sampling rate must be reduced to 47 KHz which is the Nyquist sampling rate. This is carried out in 4-stages. The first stage involves the reduction of the sampling frequency by the decimation factor of 16. This is done by the CIC filter.

The remaining 3 stages involve the reduction of the sampling frequency by the decimation factor of 2 only which are carried out by the first half band, droop correction and the second half band respectively. The Simulation results for all 4 stages are given in figure 7.

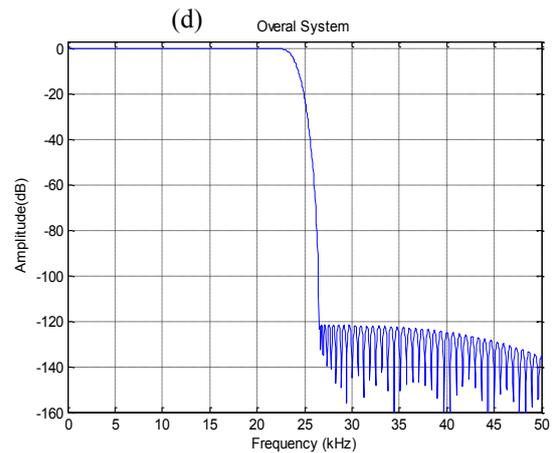

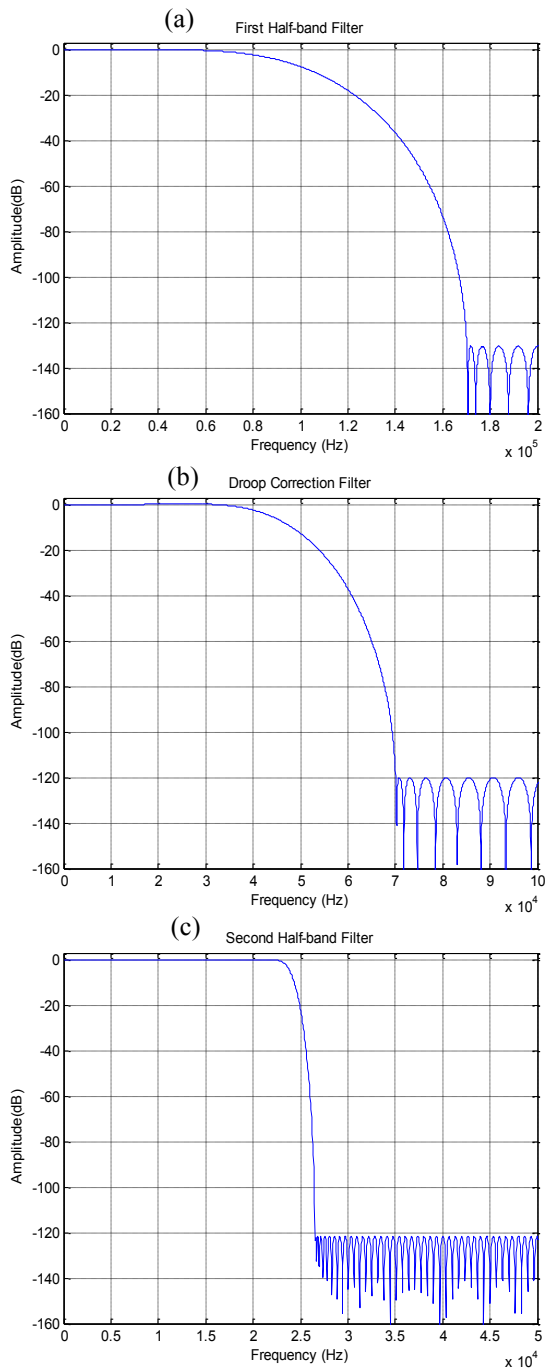

*Fig. 7* frequency response of (a) first half band filter (b) droop correction (c) second half band filter (d) overall decimation process , Over sampling ratio=128

Figure 8 shows the measured baseband output spectra before (Figure 8(a)) and after (Figure 8(b)) the decimation functions.

The CIC filter Verilog code was written and simulated by Matlab software.

It is found Signal to Noise ratio (SNR) is 141.56 dB in sigma delta modulator output and SNR is increased to 145.35 dB in the decimation stages.

To improve the signal to noise ratio, word length of recursive CIC filter should be increased but the speed of filter calculation is also decreased.

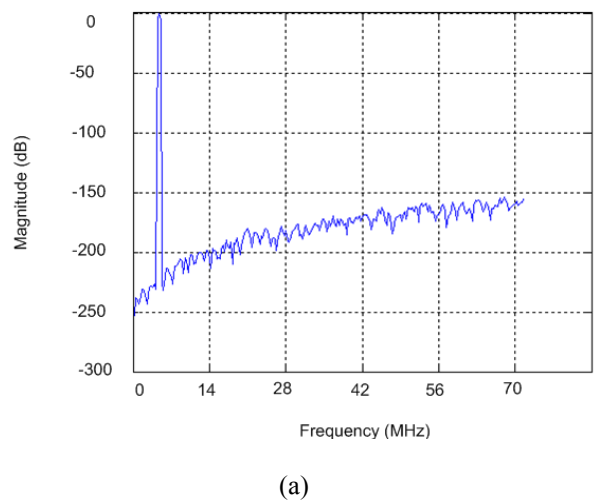

(a)

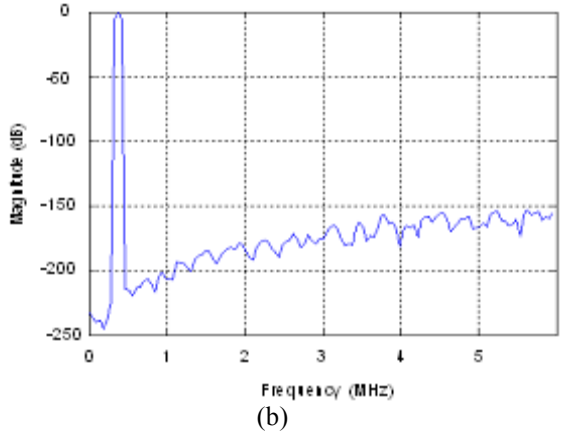
(b)

*Fig. 8* Signal spectra (a) Output sigma delta modulator SNR (b) Output CIC filter SNR

## 5 Conclusions

Recursive CIC filters have been designed and investigated. Enhanced high Speed CIC filters was obtained by the pipeline structure and by using the modified carry look-ahead adder (MCLA). The evaluation indicates that the pipelined CIC filter with MCLA adder is attractive due to high speed when both the decimation ratio and filter order are not high as stated in the Hogenauer Comb filter. Since the first stage of CIC filter require maximum word length and also because of the recursive loop in its structure, thus power is limited by the calculation in integrator stage, so the truncation will reduced the power consumption to obtain high speed operation.